\documentclass[12pt]{article}
\usepackage{setspace}
\usepackage{color}
\usepackage{amsmath, amssymb,slashed,url}
\usepackage{amsthm}
\usepackage{graphicx,epsfig}
\usepackage{cite}

\newcommand{\beq}{\begin{equation}}
\newcommand{\eeq}[1]{\label{#1}\end{equation}}
\newcommand{\bea}{\begin{eqnarray}}
\newcommand{\eea}[1]{\label{#1}\end{eqnarray}}

\def\draftnote#1{{\color{red} #1}}

\begin{document}
\begin{titlepage}
\vspace{20pt}
\begin{center}
{\huge   Soft Hair as a Soft Wig}

\vspace{18pt}

{\large  Raphael Bousso$^a$  and Massimo Porrati$^b$}

\vspace{12pt}

{$^a$\em Center for Theoretical Physics and Department of Physics, 
University of California, Berkeley, CA 94720, USA} \\ and \\ 
{\em Lawrence Berkeley National Laboratory, 
Berkeley, CA 94720, USA}\footnote{email: bousso@lbl.gov} 

\vspace{12pt}

{$^b$\em Center for Cosmology and Particle Physics, \\ Department of Physics, New York University, \\726 Broadway, 
New York, NY 10003, USA}\footnote{email: massimo.porrati@nyu.edu}

\end{center}

%
%
\begin{abstract} We consider large gauge transformations of gravity and electromagnetism in $D=4$ asymptotically flat spacetime. Already at the classical level, we identify a canonical transformation that decouples the soft variables from the hard dynamics. We find that only the soft dynamics is constrained by BMS or large $U(1)$ charge conservation. Physically this corresponds to the fact that sufficiently long-wavelength photons or gravitons that are added to the in-state will simply pass through the interaction region; they scatter trivially in their own sector. This implies in particular that the large gauge symmetries bear no relevance to the black hole information paradox. We also present the quantum version of soft decoupling. As a consistency check, we show that the apparent mixing of soft and hard modes in the original variables arises entirely from the long range field of the hard charges, which is fixed by gauge invariance and so contains no additional information.  \end{abstract}

\end{titlepage}


\section{Introduction}

Large gauge transformations in asymptotically flat spacetime are generated by an infinite set of charges, $Q[f]$~\cite{sqed1,cl,sqed2}. Here $f(\Theta)$ is an arbitrary function on the celestial sphere, so there is one independent charge per solid angle $\Theta$. Unlike a pure gauge constraint, the value of $Q[f]$ does not vanish identically, so it connects distinct states in phase space.

The charge consists of a soft and a hard part, $Q[f]=Q_s[f]+Q_h[f]$, defined in more detail below. In massless electromagnetism, $Q_s$ generates an angle-dependent change of the gauge potential, and $Q_h$ generates the compensating phase changes of charged fields. Another important example is gravity, where $Q[f]$ generates an asymptotic diffeomorphism called Bondi-van der Burg-Metzner-Sachs (BMS) supertranslation~\cite{bms}. In this case, $Q_h$ generates an angle-dependent time translation of all radiative modes, and $Q_s$ generates a compensating deformation of asymptotic coordinate spheres~\cite{sbms1,sbms2,clbms}. 

The soft and hard parts are not separately conserved. But with a suitable identification of the gauge potentials near spatial infinity, the total charge is conserved: $Q^+[f] = Q^-[f]$. (Superscripts $\pm$ refer to past and future null infinity, ${\cal I}^\pm$.) 

Thus, it would appear that asymptotic symmetries place interesting constraints on the scattering problem, at least in $D=4$ dimensions. Indeed, Hawking, Perry and Strominger~\cite{hps1,hps2} have speculated that the conservation laws associated with BMS supertranslations or large $U(1)$ transformations may have some bearing on the black hole information paradox~\cite{hawk}.

The purpose of this note is to point out that this is not the case. We show that the asymptotic symmetry group is entirely accounted for by the physically obvious freedom to add sufficiently soft (long-wavelength) particles, which propagate freely through the hard interaction region. It is clear that soft particles cannot affect hard scattering, for otherwise experiments at the LHC could not be analyzed without detailed knowledge of the cosmic microwave background. 

The simplicity of the soft dynamics is somewhat obscured in $D=4$, because the soft sector inherits some nontrivial dynamics from the hard scattering. Hard particles have a long range field, and in $D=4$ this tail corresponds to excited radiative modes at arbitrarily large wavelength. (This is not the case in $D>4$.) When hard particles scatter, their tails stay attached. However, this is the {\em only} nontrivial dynamics in the soft sector. Since it is completely determined by the hard process via the gauge constraints, it yields no independent information about the hard scattering. 

To make this explicit, we perform a simple canonical transformation, or ``dressing,'' on the dynamical variables. The transformation removes the dynamical effects of the hard scattering on the soft sector. We find that the dynamics of the two sectors factorizes. Our result is closely related to the IR factorization used in~\cite{mp} to remove soft hair (see also~\cite{gabai,averin,gomez,panchenko,shiu}). Ref.~\cite{mp} uses quantum factorization formulas that apply to photons or gravitons with small but nonzero frequency; whereas the factorization defined here applies directly to the zero-frequency modes that are at the origin of the infinite degeneracy of vacua. We also find that it applies both at the classical and at the quantum level.

We also show that the BMS and $U(1)$ large gauge charge {\em does} vanish, like a pure gauge transformation, for a class of dressed states. This class is general enough that an arbitrary state can be reached by adding further soft excitations that are unrelated to the long range field of the hard particles. The fact that $Q[f]$ need not vanish can thus be understood entirely in terms of our freedom to add these additional unconstrained soft modes, which scatter trivially.

In future work we will argue that with standard definitions of $Q_s$ and its conjugate, neither quantity is observable. However, this is an orthogonal problem, and it can be remedied~\cite{bp-prep}. In the present paper, we consider a different question, namely whether large gauge symmetries nontrivially constrain the scattering problem in $D=4$. The answer is no, regardless of whether the soft sector is observable.

In Section 2, we introduce the asymptotic symmetries and associated conserved charges. Section 3 is the heart of the paper. In Sec.~3.1 we consider the classical case. We perform a canonical transformation and show that the evolution of the new, dressed variables is {\em completely independent} of the soft-hair state. The soft sector decouples from the hard dynamics and becomes trivial. 
In Sec.~3.2, we obtain the quantum mechanical counterpart of this result. This is related to the well known infrared factorization property, which guarantees the absence of soft infrared divergences in QED or perturbative quantum gravity. In Section 4 we derive the physical interpretation of the dressing transformation. We show that it subtracts from the soft sector the excitations caused by the long range field of the hard particles. Thus, it removes its only nontrivial dynamics.
Appendix A shows how to dress gravitational memories defined by the permanent displacement of massive detectors subject to gravitational radiation in a way that makes factorization of the BMS supertranslation dynamics manifest. 

We need not include a discussion of the ``black hole horizon charges'' considered in~\cite{hps1,hps2}. In any complete quantum treatment,\footnote{Even when the semiclassical approximation is valid, the definition of ``horizon charges'' is subject to severe ambiguities, and we are not aware of a controlled limit where they make sense. If one fixes the radius of a Schwarzschild black hole while taking $G\to\infty$, then the total hard flux at ${\cal I}^+$ diverges because of
the Hawking radiation~\cite{hawk}. Moreover, the evaporation timescale diverges, so it is not clear what one would mean by soft modes and charges.} the only well-defined observables reside at ${\cal I}^\pm$, and our analysis applies directly to those observables. 

\section{Asymptotic Gauge Transformations}

In this section, we introduce the large gauge transformations of electromagnetism and gravity in four dimensions. We 
identify associated charges and conservation laws. We will use the formalism developed in~\cite{ashtekar}
and extensively used in~\cite{sqed1,sqed2,sbms1,sbms2,hps1,hps2}.

\subsection{Large $U(1)$ Symmetry}

Large $U(1)$ symmetries are Abelian gauge transformations that do not vanish at null infinity ${\cal I}^\pm$ of Minkowski space.

\subsubsection*{Transformation at ${\cal I}^+$}

The large $U(1)$ symmetry is generated by a charge~\cite{sqed1,sqed2} defined at $\mathcal{I}^+_-$, 
the past boundary of $\mathcal{I}^+$
\beq
Q^+[f]={1\over e^2}\int_{{\cal I}^+_-} f *F = {1\over e^2} \int_{{\cal I}^+} df \wedge *F + \int_{{\cal I}^+} f *J - \int_{{\cal I}^+_+} f *F ~.
\eeq{m1}
The last equality uses Maxwell's equations $d*F=e^2 *J$. We assume that there are massless charges that can reach ${\cal I}^+$. We also assume that no charge remains inside the spacetime at late times, so the last term, defined at the future
boundary of $\mathcal{I}^+$, is equal to zero.

In retarded coordinates 
\beq
ds^2= -du^2 -2du\,dr +r^2 \gamma_{AB}\,d\Theta^A d\Theta^B~,
\eeq{m1a}
the charges then take the form
\bea
Q^+[f] &=& Q^+_h + Q^+_s~,\nonumber \\
Q^+_h [f]&=& \int_{{\cal I}^+} du\, d^2\Theta \sqrt{\gamma}\, f j_u~, \qquad Q^+_s[f]={1\over e^2} 
\int _{{\cal I}^+} du\, d^2\Theta \sqrt{\gamma}\, f D^A F_{u A} ~.
\eea{m2}
Here the covariant derivative $D_A$ is defined with respect to the metric
$\gamma_{AB}$ on the unit coordinate sphere, by $D_A  \gamma_{BC}=0$. Indices $A,B$ are raised and lowered with $\gamma_{AB}$. 

The large gauge trans\-formation preserves the gauge $A_r=0,A_u|_{{\cal I}^+}=0$; so the soft charge $Q_s[f]$ can also be written as 
\begin{equation}
Q_s[f]=e^{-2} \int _{-\infty}^{\infty} du \int d^2\Theta \sqrt{\gamma}\, f D^A \partial_u A_A~.  
\label{eq-qsu1}
\end{equation}
The hard charge $Q^+_h$ is defined in terms of the boundary current $j_u\equiv\lim_{r\rightarrow\infty} r^2 J_u$. Physically this corresponds to massless charged particles that reach ${\cal I}^\pm$.  

The soft charge $Q^+_s[f]$ is the ``electromagnetic memory'' of an infinite time interval. Physically this is not even approximately observable~\cite{bp-prep}. However, a finite-time memory can be generated by soft photons. It can be observed as a change in the relative phases of test charges stationed at large radius. An observable version of $Q_s[f]$ will be introduced in~\cite{bp-prep}. The distinction is irrelevant for our present purposes, so we will use the idealization (\ref{eq-qsu1}).

$Q_s[f]$ has vanishing Poisson brackets\footnote{Once Gauss law constraints are solved and independent dynamical variables have been identified, there is no difference between the Poisson brackets of those variables and the Dirac bracket.} with all matter fields and with the radiative modes $F_{uA}|_{{\cal I}^+}$.  To obtain a symplectic structure in the soft sector, one can introduce a mode $\phi^+(\Theta)$ by~\cite{sqed1}
\beq
\lim_{u\rightarrow -\infty}A_A(u,r,\Theta)|_{{\cal I}^+}=-\partial_A \phi^+(\Theta).
\eeq{m3}
and impose the Poisson bracket
\begin{equation}
[Q^+_s[f],\phi^+(\Theta)]=-i f(\Theta).
\end{equation}
Thus
\begin{equation}
\Pi^+(\bar{\Theta})\equiv Q^+_s[\delta^2(\Theta -\bar{\Theta})]
\end{equation}
is the momentum canonically conjugate to $\phi^+(\Theta)$. 

As defined, $\phi^+$, like $Q_s$, is completely unobservable. Again, an alternate definition can be given in terms of observable soft photon wavepackets~\cite{bp-prep}. This does not affect our analysis here, so we adopt the conventional definition above.

Expansion into spherical harmonics, $f(\Theta)=\sum_{l=0}^\infty\sum_{m=-l}^{l} f_{lm} Y_{lm}(\Theta)$, yields a countable infinity of canonically conjugate operators.  Upon quantization they become creation and annihilation operators, which generate a ``soft'' Fock space with a countable basis.

\subsubsection*{Transformation at ${\cal I}^-$ and Matching}

Near ${\cal I}^-$ the metric is $ds^2= -dv^2 +2dvdr +r^2 \gamma_{AB}d\Theta^A d\Theta^B$. 
The angular coordinates on ${\cal I}^-$ are identified with those at ${\cal I}^+$ by requiring that a null geodesic issuing from 
${\cal I}^-$ and passing through  $r=0$ ends up at the same value of the angular coordinate. In other words, the 
angles $\Theta$ here are {\em antipodally} identified with those in Eq.~(\ref{m1a}). 
The conserved charge is 
\bea
Q^-[f] &=& Q^-_h + Q^-_s~, \\
Q^-_h [f]&=& \int_{{\cal I}^-} dv\, d^2\Theta \sqrt{\gamma}\, f j_v~, \qquad Q^-_s[f]={1\over e^2} 
\int _{{\cal I}^-} dv\, d^2\Theta \sqrt{\gamma}\, f D^A F_{v A}~.
\eea{m5}
The mode canonically conjugate to $\Pi^-(\bar{\Theta})\equiv Q^{-}_s[f]$, $f(\Theta)= \delta^2(\Theta -\bar{\Theta})$, is $\phi^-(\Theta)$, 
which is defined similarly to $\phi^+$ as
\beq
\lim_{v\rightarrow +\infty}A_A(v,r,\Theta)|_{{\cal I}^-}=-\partial_A \phi^-(\Theta).
\eeq{m6}

The soft degrees of freedom $\phi^+,\Pi^+$ on ${\cal I}^+$ are identified with those on ${\cal I}^-$ by imposing 
matching conditions. This is a necessary step to define a scattering problem.

The first condition is conservation of the total charge, $Q^+[f] = Q^-[f]$.  The rationale for imposing this condition comes from studying the electromagnetic fields produced by moving charges that interact in a finite region of Minkowski space. In physically sensible generic solutions~\cite{sqed2,stro-rev} the field strength $F_{ru}$ at any angle $\Theta$ on $ {\cal I}^+_-$ equals $F_{rv}$ on ${\cal I}^-_+$ at the antipodal angle.  By Eq.~(\ref{m1}), the total charge can be expressed only in terms of those quantities and so is antipodally conserved at every angle. Expressed as the sum of soft and hard charge, the matching condition becomes
\beq 
\Pi^+(\Theta) +\int_{{\cal I}^+}du \sqrt{\gamma}\,j_u(u,\Theta) =\Pi^-(\Theta) + \int_{{\cal I}^-} dv \sqrt{\gamma}\,j_v(v,\Theta).  
\eeq{m9} 
This matching condition can be violated by superimposing free gauge fields that do not vanish at $v\rightarrow +\infty$ to initial field configurations that do satisfy it. Finite-energy free fields do vanish at $v=\infty$, so Eq.~(\ref{m9}) may turn out to be equivalent to imposing finite energy, though we are not aware of a proof. 

The second matching condition can be chosen to be 
\beq 
\phi^+(\Theta)=\phi^-(\Theta).  
\eeq{m8} 
This choice is invariant under CPT and Lorentz transformations and is consistent with the asymptotic behavior of generic potentials for moving charges~\cite{sqed2,stro-rev}. Matching conditions~(\ref{m9},\ref{m8}) lead to standard soft photon/graviton theorems in perturbative quantum field theory~\cite{sqed1,sbms1}. (However, we are not aware of a relation of (\ref{m8}) to first principles such as boundedness of the energy.)

\begin{quotation}
  The matching conditions~(\ref{m9},\ref{m8}) are the essential ingredient for deriving our factorization result in Section 3. 
  The triviality of soft hair is a consequence of Eqs.~(\ref{m9},\ref{m8}) and {\em only\/} of them. No input about the 
  dynamics of the interior of spacetime is required to derive the results of Section 3.
\end{quotation}
  
  
  \subsection{BMS Supertranslations}

The metric of an asymptotically flat space near  ${\cal I}^+$ is (see~\cite{hps2} for notation and normalizations)
 \beq
 ds^2= -du^2 - 2dudr + r^2 \gamma_{AB} d\Theta^A d\Theta^B 
+ rC_{AB} d\Theta^A d\Theta^B +...
 \eeq{m17}
The Bondi news is $N_{AB}=\partial_u C_{AB}$. 

The metric retains the above form under large diffeomorphisms generated by the BMS charge~\cite{sbms1,sbms2,hps2}
 \bea
Q^+&=& Q^+_h+Q^+_s, \nonumber \\
Q_h[f] &=& {1\over 4\pi} \int_{{\cal I}^+} du\, d^2\Theta \sqrt{\gamma}\, f(\Theta) T_{uu} ,\qquad 
 Q^+_s[f]=-{1\over 16\pi G} \int_{{\cal I}^+} du\,d^2\Theta \sqrt{\gamma}\, f(\Theta) D^A D^B N_{AB}, \nonumber \\
 T_{uu}&=&{1\over 8G}N_{AB}N^{AB} +
 \lim_{r\rightarrow\infty} r^2 T^M_{uu}, \qquad ~~~~~~T^M=\mbox{matter stress-energy tensor}.
 \eea{m19}
$Q^+_s$ commutes with $N_{AB}$ and matter fields. To complete the symplectic structure, we introduce a boundary field $C(\Theta)$ via
\beq
\lim_{u\rightarrow -\infty} C_{AB}(u,\Theta) = -2D_A D_B C(\Theta) + \gamma_{AB} D^2 C(\Theta)~;
\eeq{m18}
and we impose the Poisson bracket
\begin{equation}
[Q^+_s[f],C^+(\Theta)]=-i f(\Theta)~.
\end{equation}
Then 
\begin{equation}
\Pi^+(\bar{\Theta})\equiv Q^+_s[\delta^2(\Theta-\bar{\Theta})]
\end{equation}
is the momentum canonically conjugate to $C^+(\Theta)$.

One can likewise define an asymptotic metric near ${\cal I}^-$ in terms of angles $\theta^A$, radius $r$ and advanced time $v$:
\beq
 ds^2= -dv^2 - 2dvdr + r^2 \gamma_{AB} d\theta^A d\theta^B 
+ rC^-_{AB} d\theta^A d\theta^B +....
\eeq{m19a} 
The past Bondi news is $N^-_{AB}=\partial_v C^-_{AB}$. The past charges $Q^-[f]=Q_h^-[f]+Q^-_s[f]$ and the canonical 
pairs $C^-(\theta)$ and $\Pi^-(\theta)$ at ${\cal I}^-$ are defined in complete analogy with the future charges 
at ${\cal I}^+$.
 
The matching of angular coordinates on ${\cal I}^\pm$ can be performed by continuing the generators of ${\cal I}^+$ through space-like infinity $i_0$. This again corresponds to an antipodal identification of angles, to the extent that this bulk notion is well-defined (e.g., in the vacuum).  After this identification we drop the distinction between angular coordinates on ${\cal I}^+$ and ${\cal I}^-$ and we call them both $\Theta^A$, where it is understood that the same label corresponds to antipodal generators on ${\cal I}^\pm$.

We therefore get the following matching for the soft degrees of freedom:
 \bea
 \Pi^+(\Theta)+ \eta^+(\Theta) &=& \Pi^-(\Theta)+ \eta^-(\Theta), \qquad C^+(\Theta)=C^-(\Theta), \nonumber \\
 \eta^+(\Theta) &\equiv & {1\over 4\pi} \int_{{\cal I}^+} du \sqrt{\gamma}\, T_{uu} (u,\Theta), \qquad
 \eta^- (\Theta) \equiv  {1\over 4\pi} \int_{{\cal I}^-} dv 
 \sqrt{\gamma}\, T_{vv} (v,\Theta)~.
 \eea{m21}
 Apart from obvious  changes of names these are the same conditions that we encountered in the $U(1)$ problem.  
 
\section{Factorization}

\subsection{Classical Factorization}

The relation between past {\em in} variables at ${\cal I}^-$ and future {\em out} variables at ${\cal I}^+$ defines a classical 
scattering problem. We can solve this problem directly in the BMS case without passing through the ``warmup" case of 
large $U(1)$ symmetry, since the formalism is identical for both cases.  

In classical mechanics the relation between past and future is given by a symplectic transformation, which is itself defined by a generating functional~\cite{arn}. The generating functional can be taken to depend on the initial coordinates $C^-,\ldots $ and final momenta $\Pi^+,\ldots\, $, where $\ldots$ stands for all other canonical coordinates.

To define a generating functional one must also partition the hard Bondi news into symplectic sets of commuting variables. This can be done in many ways. A convenient one, tailored to make contact with the definition of creation and annihilation operators for gravitons, is to define the positive-frequency part of the news $N_{AB} (u, \Theta)$ as ``coordinates'' and the negative-frequency ones as ``momenta.'' (These variables are independent of the boundary variables $C^\pm$, unlike the $C_{AB}^\pm$.)

The form of the matching conditions~(\ref{m21}) suggests that we perform the following canonical transformation on the {\em in} and {\em out} variables:
 \bea
\Pi^+(\Theta) \rightarrow \Pi^{+D}(\Theta) &=&  \Pi^+(\Theta) + \eta^+(\Theta)~, \nonumber \\
 N_{AB} (u,\Theta) \rightarrow N^{D}_{AB}(u,\Theta) &=& N_{AB} (u - C^+(\Theta),\Theta)~, \nonumber \\
 \Pi^-(\Theta) \rightarrow \Pi^{-D}(\Theta) &=& \Pi^-(\Theta)+\eta^-(\Theta)~, \nonumber \\
 N^{-}_{AB} (v,\Theta) \rightarrow N^{- D}_{AB}(v,\Theta)&=&N^{-}_{AB} (v-C^-(\Theta),\Theta)~. 
 \eea{mp2}
The new {\em out} variable $\Pi^{+D}$ has canonical commutation relations with $C^+$ and $N^{D}_{AB}$; in particular, it {\em commutes} with $N_{AB}^D$. The {\em in} variable $\Pi^{-D}$ enjoys the same commutation relations with the transformed canonical variables defined on ${\cal I}^-$. 
 
In terms of the new variables the matching conditions are 
 \beq
C^+(\Theta) = C^-(\Theta)~, \qquad \Pi^{+D}(\Theta) = \Pi^{-D}(\Theta) ~.
\eeq{mp3} The generating functional of asymptotic time evolution is a function $F[C^-,N^{-D}_{+AB}, \Pi^{+D}, N^{D}_{-AB}]$ of initial coordinates and final momenta.\footnote{$F$ depends also on additional symplectic variables associated to matter, when matter fields are present. They are just additional hard variables conceptually identical to the dressed Bondi news.} Here the subscripts $+$ and $-$ denote positive and negative frequencies in $u$ and $v$.

The matching conditions imply
\beq
C^-(\Theta^A) = {\delta F \over \delta \Pi^{+D}(\Theta^A)}~ , \qquad \Pi^{+D}(\Theta^A)=  
{\delta F \over \delta C^-(\Theta^A)}~.
\eeq{mp4}
This equation is easily solved by
\beq
F=\int d^2\Theta~ C^-\,\Pi^{+D} + f[N^{-D}_{+AB}, N^{D}_{-AB}]~,
\eeq{mp5}
where $f$ is {\em independent of} $C^-, \Pi^{+D}$. 

\begin{quotation}
This is a key result of our  paper. It shows that BMS symmetry does not constrain the dynamics of the hard degrees of freedom at all, because it tells us nothing about the functional $f$.\footnote{The memories $\Pi^\pm(\Theta)$ can be defined in terms of the permanent displacement of sets of massive particles moving along appropriate world-lines. Appendix A shows how to dress such quantities in a manner that makes the factorization of hard dynamics explicit.}
\end{quotation}

To be concrete, suppose that we are given an initial configuration described by the canonical variables 
$C^-,\Pi^-,N^{-}_{AB}$. Naively, we might think that the infinite number of conserved BMS charges, together with the 
intricate way in which hard and soft modes mix during time evolution, could imprint some information about the initial state 
on the final configuration of the soft modes $C^+,\Pi^+$. Equation~(\ref{mp5}) shows that this is a mistake.

Once the initial state is rewritten in terms of properly ``dressed'' variables, the evolution of the soft modes decouples 
completely from that of the hard modes. In fact, soft modes evolve trivially.  Even more importantly, the evolution of the hard 
states is {\em the same} for any configuration of soft states. 

The apparent nontrivial mixing of hard and soft modes under time evolution is just an illusion due to a bad choice of 
coordinates.  In other words, the soft hair is a wig. It can be pulled off without affecting the rest of the dynamics. Its ``hairs" 
are conserved because they decouple, not because they carry any information about the rest of the world.\footnote{The soft 
particles do carry their own information, of course. This information is independent of the hard scattering data, and it cannot
 be accessed~\cite{Bou16,BouHal16,Bousso:2016beo} on the shorter timescales sufficient for producing and measuring the hard in and out 
 states.}

We have proven factorization of the soft dynamics in classical mechanics, but an analogous result holds also in quantum field theory for the S-matrix, as we will show in the next subsection.

\subsection{Quantum Factorization}

We do not have a complete quantum theory of gravity in asymptotically flat space-times; therefore, any result that we may hope to obtain must rely on additional assumptions. Ours will be that there exists a unitary S-matrix that maps quantum fields defined on ${\cal I}^-$ to fields defined on ${\cal I}^+$. 

With this assumption, we need not investigate the effect of horizons,
which are associated with ephemeral intermediate states from the point of view of the S-matrix. 
We will now describe the quantum analog of the classical ``dressing" of canonical variables given in Eq.~(\ref{mp2}), as 
well as the analog of the canonical transformation defined by the generating functional in Eq.~(\ref{mp5}).

A classical canonical transformation becomes a unitary transformation in quantum mechanics. Its form is almost uniquely dictated, up to a state-independent phase, by its action on coordinates and momenta, given in Eq.~(\ref{mp2}). The previous two sentences should be qualified in field theory, because the presence of infinitely many degrees of freedom may make the transformation only formally unitary (that is, unitary in a Hilbert space larger than the physical one). This is a good thing, because otherwise we would find that formally unitarily equivalent descriptions of scattering give inequivalent results.\footnote{The ``dressing'' operator that we shall find is closely related to that used in~\cite{chung,kibble,fk} to tame the infrared problem of QED. While formally unitary, the latter maps any vector belonging to the Fock space of photons, $H$, into a vector orthogonal to {\em all} the vectors in $H$ (see e.g.\ Eq.~12 of~\cite{fk}). This is of course incompatible with the dressing being a well-defined unitary operator in $H$.}  With this caveat in mind, let us proceed to a formal definition of the unitary operator that implements the canonical transformation~(\ref{mp2}) on states.

The classical mapping between {\em in} and {\em out} variables translates naturally into the Heisenberg picture of quantum evolution, where operators evolve while states do not. Heisenberg-picture operators will carry a subscript $H$. We need two operators\footnote{We apologize for deviating from our convention of using $\pm$ superscripts associated with ${\cal I}^\pm$. The consistent notation $V^+_H, V^-_H$ would be cumbersome because we frequently refer to the inverse, $(V^\pm)^{-1}$.} $U_H,V_H $ that transform the {\em in} and {\em out} soft variables as
\beq
U_H \Pi^+_H U^{-1}_H = \Pi^+_H + \eta^+_H, \qquad V_H \Pi_H^- V_H^{-1} = \Pi^-_H + \eta^-_H~.
\eeq{mp6}
Since the canonical commutator is
 \beq
 [\Pi^+_H(\Theta),C^+_H(\Theta')] = -i \delta^2(\Theta-\Theta')~,
 \eeq{mp7}  
 the operators $U_H,V_H$ are
\beq
U_H= e^{-i\int d^2\Theta\, C_H^+(\Theta) \eta_H^+(\Theta)}~,\qquad 
V_H= e^{-i\int d^2\Theta\, C_H^-(\Theta) \eta_H^-(\Theta)}~,
 \eeq{mp8}
up to a c-number phase.
  
As we mentioned earlier, the integral over $\Theta$ can be replaced with a sum over angular momenta by expanding $\eta_H^+(\Omega),C_H^+(\Omega)$, 
 $\eta_H^-(\Omega),C_H^-(\Omega)$ in spherical harmonics. 
 The quantum analog of the matching conditions~(\ref{m21}) is
 \beq
 U_H \Pi^+_H U^{-1}_H =V_H \Pi_H^- V_H^{-1}~, \qquad  U_H C^+_H U^{-1}_H =V_H C_H^- V_H^{-1}~.
 \eeq{mp9}
 On the other hand, the {\em in} and {\em out} Heisenberg operators are mapped into each other, by our assumption, by a unitary evolution operator $S_H$:
 \beq
 \Pi^+_H=S_H^{-1} \Pi_H^-S_H~, \qquad C^+_H= S_H^{-1} C^-_H S_H~.
 \eeq{mp10}
 Eqs.~(\ref{mp9},\ref{mp10}) imply that the matrix $S_HU_H^{-1}V_H$ commutes with 
 $\Pi^-_H,C^-_H$; but the
 latter are canonically conjugated operators that commute with all other dynamical variables. Thus, by Schur's lemma, 
 $S_HU_H^{-1}V_H$ is
 proportional to the identity on any irreducible representation of the soft canonical commutator algebra. $S_HU_H^{-1}V_H$
 acts as a nontrivial unitary matrix
 only on the Hilbert space associated to the ``hard" operators $N_{AB}$ (plus eventual matter fields). 

This property allows us to introduce a factorized S-matrix $\hat{S}_H$ that does not act on the Hilbert space of soft modes. By their definition, the dressing operators~(\ref{mp9}) obey $U_H= S^{-1}_H V_H S_H$; it follows that
 \beq
 \hat{S}_H = S_H U_H^{-1} V_H= V_H^{-1} S_H V_H \rightarrow S_H = V_H \hat{S}_H V_H^{-1}~.
 \eeq{mp11}
This is almost a factorization formula for the S-matrix. 

The last step we need is to recall that the usual S-matrix, that is the one that is computed by Feynman diagrams in perturbation theory, is defined in the interaction representation, where operators evolve with a ``free'' S-matrix $S_0$, while states evolve with $S\equiv S_0^{-1}S_H$. Operators in interaction representation will carry no subscript. A standard choice is that  operators in Heisenberg and interaction representation coincide on
 ${\cal I}^-$. In the interaction representation, the {\em out} operators 
 on ${\cal I}^+$ are defined as $O_{out}=S_0^{-1} O_{in}S_0$, so $U\equiv S_0^{-1}VS_0$.  
 Finally, equation~(\ref{mp11}) becomes
 \beq
 S=U \hat{S} V^{-1}~, \qquad \hat{S}\equiv S_0^{-1} \hat{S}_H~.
\eeq{mp12}

Readers familiar with the problem of infrared divergences in quantum field theory will recognize the similarity between the factorization formula~(\ref{mp12}), and the Block-Nordsieck factorization that is the key to solving the ``infrared catastrophe." This is not a coincidence: Eq.~(\ref{mp12}) is the extreme infrared limit of the formulas used in~\cite{mp} to factorize soft $U(1)$ and BMS hair. The finite-frequency counterparts of $U,V$ were introduced in 1965 for QED~\cite{chung}. They were further studied in~\cite{kibble,fk}. As usual in theoretical physics, they are known by the name of the last authors that discovered them.\footnote{The relation between various formulations of the QED dressing factors was recently revisited in~\cite{kprs}.}

\subsubsection*{Factorization of Large $U(1)$ Soft Hair Dynamics}
 
Classical and quantum factorization formulas in this case 
are completely analogous to the BMS case after the obvious substitution
\beq
C^+\rightarrow \phi^+~, \qquad C^-\rightarrow \phi^- ~.
\eeq{m9a}
A field on ${\cal I}^+$ carrying no  $U(1)$ charge needs no dressing, since it already commutes with the charge $Q^+[f]$,
while a field $\Psi$ of charge $q$  commutes with $Q^+[f]$ after the dressing
\beq
 \Psi(u,\Theta) \rightarrow e^{-iq\phi^+(\Theta)}\Psi(u,\Theta)~.
 \eeq{m9b}
 
\section{Classical Dressing}


We defined dressed variables somewhat formally through the canonical transformations~(\ref{mp2}), (\ref{m9a}), or their quantum analogues, the dressing operators $U,V$. In this section, we will give a physical interpretation of this operation: it removes the contribution to the soft charge coming from the long-range field of the hard particles. We do this by explicitly computing this contribution and showing that it equals minus the hard charge, up to a constant.

\subsection{$U(1)$ Classical Dressed States}

In the 
metric~(\ref{m1a}) the angular components of Maxwell's equations are
\beq
-\partial_r (\sqrt\gamma \gamma^{AB}F_{uB}) - \partial_u (\sqrt\gamma \gamma^{AB}F_{rB}) +
\partial_r (\sqrt\gamma \gamma^{AB}F_{rB}) +{1\over r^2} \partial_C (\sqrt\gamma \gamma^{CD}\gamma^{AB}F_{DB}) =
 \sqrt\gamma \gamma^{AB}e^2 J_B~.
\eeq{m107}
So, by integrating in 
$r$ and using $F_{rA}={\cal O}(r^{-2})$~\cite{sqed1} we arrive at a particular solution:
\beq
F_u^A(r,u,\Theta)=- \int_0^r ds \big[e^2J^A(s,u,\Theta) +\partial_u F^A_r(r,u,\Theta)\big]+ D_C X^{CA}(r,u,\Theta)~.
\eeq{m108}
The function $X^{AB}$ is antisymmetric. Using the explicit form of the soft charge
$Q^+_s[f]$ given in Eq.~(\ref{m2}) we see that the term proportional to $X^{AB}$ vanishes after the angular 
integration; therefore,  the soft charge of solution~(\ref{m108}) is 
\beq
Q^+_s[f]=-\int_0^\infty ds \int d^2\Theta \sqrt{\gamma}\,f(\Theta)\Big[{1\over e^2}  
D_AF^A_r(s,u,\Theta)\Big|^{u=+\infty}_{u=-\infty} +\int_{-\infty}^{+\infty} du\, D_AJ^A(s,u,\Theta)\Big]~.
\eeq{m109}
The first term in brackets vanishes at each finite value of $s$, because at time-like infinity $i^\pm$ all field strengths vanish 
in massless QED. 

In the metric~(\ref{m1a}), the current conservation equation is
\beq
-\partial_r \sqrt{\gamma} r^2 J_u -\partial_u \sqrt{\gamma} r^2 J_r + \partial_r \sqrt{\gamma} r^2 J_r + 
\sqrt{\gamma} D_AJ^A=0~.
\eeq{m110}
The field strengths $F_{ur}$, $F_{Ar}$ are ${\cal O}(r^{-2})$ at large $r$~\cite{sqed1}.  
The $r$ component of Maxwell's equations is
\beq
\partial_r r^2F_{ur} + D_AF^A_r = r^2e^2J_r~.
\eeq{m110a}
Substituting the asymptotic behavior of the field strengths into Eq.~(\ref{m110a}) we find 
$\lim_{r\rightarrow\infty}r^2J_r=0$.  So we have
\beq
\lim_{r\rightarrow\infty}  r^2 J_u (r,u,\Theta)
=  \int_0^\infty ds \Big[D_A J^A(s,u,\Theta)-\partial_u  s^2 J_r(s,u,\Theta)\Big]~.
\eeq{m111}
At all finite $r$, the current component $r^2J_r$ vanishes at $u=\pm \infty$, because no current escapes from $i^\pm$. 
So  the definition of $Q^+_s$ gives finally
\beq
Q^+_s[f]= - \int_{-\infty}^{+\infty} du \int d^2\Theta \sqrt{\gamma} f(\Theta) j_u(u,\Theta)=
-Q^+_h~.
\eeq{m112}
This last equation expresses the equality that we have been looking for: 
\begin{equation}
\mbox{``hard charge}=-\mbox{gravitational memory of the dressing,''}
\end{equation}
if the ambiguity in the dressing operator is resolved by the choice 
(\ref{mp8}). 

Notice that in Eqs.~(\ref{m109},\ref{m112}), only the zero frequency mode of currents contribute (because of the integral in $u$). So, the definition of the dressed state is not unique. For instance, it can be changed by adding to the $F_u^A$ defined in Eq.~(\ref{m108}) an arbitrary solution of the homogeneous Maxwell equations, $F^{\mathrm{hom}\,A}_u$, as long as its Fourier tranform $\tilde{F}^{\mathrm{hom}\, A}_u (\omega) =\int_{-\infty}^{+\infty} du \exp(i\omega u) F_u^{\mathrm{hom}\, A}$ vanishes at $\omega=0$. One can also make the total charge nonzero by adding to the soft charge only photons with frequency below any arbitrarily small infrared cutoff $\omega_0>0$.

\subsection{BMS Classical Dressing}

The equations of motion of general relativity are nonlinear, so one cannot simply define dressed states by subtracting a particular solution of Einstein's equations from the bare fields. Moreover, the definition of finite-time or finite-radius fields is inherently ambiguous. Therefore the methods of the previous section cannot be used.  The alternative method we employ here is less directly connected with the scattering problem, but it has two merits: 1) it only uses asymptotic data; 2) it makes clear that the cancellation of the hard charge can be achieved by changing only soft modes.

First of all, change the Fourier transform of the Bondi news as follows
 \beq
 \tilde{N}_{AB}(\omega) \rightarrow \tilde{N}_{AB}(\omega) + \chi(\omega)[ -2D_A D_B X(\Theta) + \gamma_{AB} D^2 
 X(\Theta)], \qquad \tilde{N}_{AB}=\int du e^{i\omega u} N_{AB}(u),
 \eeq{m115}
 where $\chi(\omega)$ is a step function obeying $\chi(\omega)=1$ for $|\omega |<\omega_0$ and $\chi( \omega)=0$
 for $|\omega |>\omega_0$. The IR cutoff $\omega_0$ can be arbitrarily small. Write the BMS charge in terms of $\tilde{N}$ 
 \bea
Q^+&=& Q^+_h+Q^+_s, \qquad Q_h[f]={1\over 4\pi G} \int {d\omega \over 2\pi} \int d^2\Theta \sqrt{\gamma}\, f(\Theta) 
\tilde{N}_{AB}\tilde{N}^{AB} + Q^+_M,\nonumber \\ 
\qquad 
 Q^+_s[f]&=&-{1\over 16\pi G} \lim_{\omega\rightarrow 0} 
 \int d^2\Theta \sqrt{\gamma}\, f(\Theta) D^A D^B \tilde{N}_{AB}(\omega), \nonumber \\
 Q^+_M &=& {1\over 4\pi} \lim_{r\rightarrow\infty}\int du\, d^2\Theta \sqrt{\gamma}\, f(\Theta) r^2 T^M_{uu}.
 \eea{m116}
 
Under the transformation~(\ref{m115}) the charges change as follows
 \beq
 Q^+_s[f] \rightarrow Q^+_s[f] +  {1\over 16\pi G} \int d^2\Theta \sqrt{\gamma} f(\Theta) D^4 X(\Theta) , \qquad 
 Q^+_h[f] \rightarrow Q^+_h[f] + {\cal O}(\omega_0).
\eeq{m117}
For $\omega_0 \rightarrow 0$ the change to $Q^+_h$ is negligible. In the same limit, the hard charge can be canceled by
satisfying the equation
\beq
D^2(D^2+2) X = -4 \int {d\omega \over 2\pi}   \tilde{N}_{AB}\tilde{N}^{AB}-16\pi G Q^+_M.
\eeq{m118}
Notice that the Bondi news is unchanged for frequencies larger than $\omega_0$.

 \subsection*{Acknowledgments} 
 It is a pleasure to thank V.~Chandrasekharan, E.~Flanagan, I.~Halpern, S.~Leichenauer, A.~Strominger, and A.~Wall for discussions.  R.B.\ was supported in part by the Berkeley Center for Theoretical Physics, by the National Science Foundation (award numbers PHY-1521446, PHY-1316783), by FQXi, and by the US Department of Energy under contract DE-AC02-05CH11231.  M.P.\ was supported in part by NSF grants PHY-1316452, PHY-1620039. M.P.\ thanks the Galileo Galilei Institute for Theoretical Physics (GGI) for hospitality and INFN for partial support during the completion of this work, within the program ``New Developments in AdS$_3$/CFT$_2$ Holography.''

\setcounter{section}{0}
\renewcommand{\thesection}{\Alph{section}}
\section{Massive Particles as Gravitational Memory Detectors}
\setcounter{equation}{0} \renewcommand{\theequation}{A.\arabic{equation}}

There is one more class of observables that has played a prominent role in the study of BMS symmetry. These are the gravitational memories measured by displacement of massive particles either moving along geodesics or kept at fixed radius and fixed angular coordinates~\cite{sgm}. These quantities define a particular, concrete procedure to measure the memories $\Pi^\pm(\Theta)$ so they cannot change the conclusions reached in the paper.  Once properly dressed, they too will manifestly factor out of the nonzero-frequency dynamics. This appendix shows explicitly how to dress these variables, but we begin by defining them.

\subsubsection*{Gravitational Memories as Scattering Data}

At large $r$ the metric in retarded coordinates is  given in Eq.~(\ref{m17}), while in advanced coordinates it is given
in Eq.~(\ref{m19a}). 
 At large but finite $r$ the antipodal map 
between retarded angular coordinates $\Theta^A$ and advanced angular coordinates $\theta^A$
 is 
 \beq
 \Theta^A= \theta^A + {1\over r}f^A(\theta,v) + {\cal O}(r^{-2}).
\eeq{m122}

Consider next a bunch of detectors at fixed positions $r$, $\Theta_i$ and retarded time $u$. As in~\cite{sgm} we 
consider nearby detectors so that the distance $L_{ij}$ between any two of them at equal time is approximately
\beq
L_{ij}^2= [r^2 \gamma_{AB} + rC_{AB}](\Theta^A_i-\Theta^A_j) (\Theta^B_i-\Theta^B_j)  .
\eeq{m123}
As $C_{AB}$ evolves from $u=-\infty$ to $u=+\infty$ the detectors' relative position changes as~\cite{sgm} 
\beq
L_{ij}|_{u=+\infty} -L_{ij}|_{u=-\infty} ={1\over 2L_{ij}} r(C_{AB}|_{u=+\infty}-C_{AB}|_{u=-\infty}) (\Theta^A_i-\Theta^A_j) (\Theta^B_i-\Theta^B_j).
\eeq{m124}
This equation does not give any new information on gravitational scattering, since it simply determines the final positions of
the detectors, $L_{ij}|_{u=+\infty}$, in terms of the change in $C_{AB}$ and the initial positions $L_{ij}|_{u=-\infty}$. 
In other words Eq.~(\ref{m124}) solves the scattering problem for a particular type of massive objects, i.e. the detectors.

Notice that an analogous equation can be written in terms of the advanced coordinates and the metric components 
$C^-_{AB}$ as
 \beq
L_{ij}|_{v=+\infty} -L_{ij}|_{v=-\infty} ={1\over 2L_{ij}} r(C^-_{AB}|_{v=+\infty}-C^-_{AB}|_{v=-\infty}) (\theta^A_i-\theta^A_j) 
(\theta^B_i-\theta^B_j).
\eeq{m125}

Even though the LHS of Eqs.~(\ref{m124},\ref{m125}) is the same, 
we cannot use those equations to express the change in $C^-_{AB}$
in terms of the change of $C_{AB}$, because we do not know the functions $f^A$ in Eq.~(\ref{m122}). The best
we can do is to use the two equations to constrain $f^A(\theta,v)$. Specifically, 
\beq
2L_{ij} \gamma_{AC}\Delta f^C_{B\, j} = C_{AB}|_{u=+\infty}-C_{AB}|_{u=-\infty}-C^-_{AB}|_{v=+\infty}+C^-_{AB}|_{v=-\infty},
\eeq{m126}
where $\Delta f^A_{B\,j}=\partial_B f^A(\theta_j,+\infty) - \partial_B f^A(\theta_j,-\infty)$. 

\subsubsection*{Dressed Gravitational Memories}

The effect of soft BMS hair on the dynamics of gravitational memories can be factored out by 
defining appropriate variables that are insensitive to the IR fields. This 
must be true also for the variables defined in this appendix, which are just
a concrete realization of gravitational memories. Dressed Bondi news were defined in Eq.~(\ref{mp2}).  
Equation~(\ref{m124}) suggests how to define the  ``dressed memory'' 
 observables that we defined here:
 \beq
 L_{ij}^D\equiv L_{ij} - {1\over 2L_{ij}} rC_{AB}(\Theta^A_i-\Theta^A_j) (\Theta^B_i-\Theta^B_j).
 \eeq{m127}
 These variables remain constant under time evolution so $L_{ij}^D|_{u=+\infty}=L_{ij}^D|_{u=-\infty}$ so they are obviously 
 independent of the soft graviton state.

  \end{document}